\documentclass[reqno,10pt,centertags]{amsart}
\usepackage{amssymb,upref,graphicx,lastpage}
\usepackage[utf8]{inputenc}
\usepackage{float}
\allowdisplaybreaks \numberwithin{equation}{section}

\usepackage{graphicx,lastpage}
\makeatletter
\def\@maketitle@hook
  {\hbox to \textwidth
    {\valign{\vss##\vss\cr
      \hbox{\includegraphics[width=30mm]{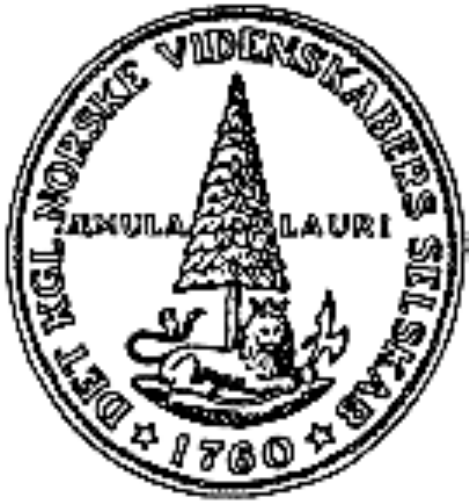}}\cr
      \noalign{\hfil}
      \hsize=0.7\hsize
      \vbox{\parindent0pt \raggedleft

         \medskip
         \sffamily
         Transactions of The\\ Royal Norwegian Society\\ of Sciences and
         Letters,\\ 
         (\textit{Trans.\ R.\ Norw.\  Soc.\  Sci.\ Lett.} 20xx(x) x--x)\par}\cr}}
  \vskip 42pt \relax}
\makeatother

\begin{document}

\title[]{Simple water-like lattice models in one dimension}

\author[]{Enrique Lomba}
\address{IQFR-CSIC, C/ Serrano 119, E-28006 Madrid, Spain}
\email{enrique.lomba@csic.es}
\urladdr{}

\dedicatory{Dedicated  to Prof, Johan S. H\o ye on occasion of his
  70th birthday.}
\thanks{}
\keywords{Phase equilibria, water-like anomalies, one-dimensional
  lattice gas, transfer matrix method, mean field.}
\date{\today}

\begin{abstract}
In this contribution we review a series of simple one dimensional
lattice models that with an appropriate choice of parameters can
account for various anomalous features of the behaviour of complex
systems such as water. In particular, we will focus on the presence of
$p-T$ fluid-solid coexistence lines with negative slope
(i.e. solids that melt upon compression), solid phases less dense than
the liquid phase, and the existence of  temperatures of maximum
density. We will see how a simple two-parameter model can reproduce 
the phase behaviour of a range of systems well known for their
anomalous behaviour regarding the temperature and pressure dependence
of properties such as density, diffusivity or viscosity. 
\end{abstract}

\maketitle

\section{Introduction}
\begin{figure}[t]
\begin{center}
\includegraphics[width=13cm,clip]{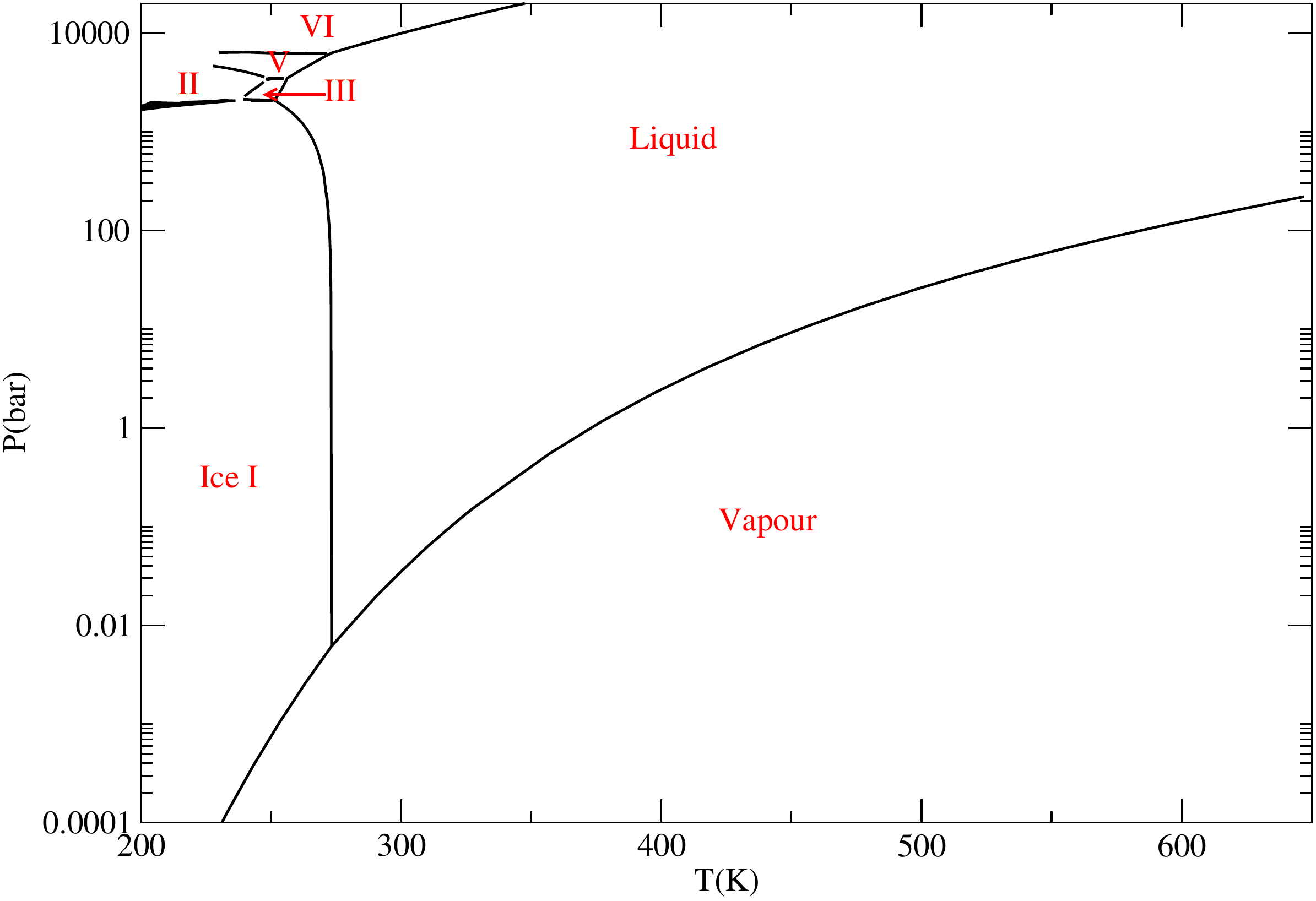}
\end{center}
\caption{Experimental phase diagram of water. Roman numerals designate
  different ice phases. Data taken from Refs.\cite{Linstrom2011,Wexler1977,Mishima1980}.}
\label{PDH2O}
\end{figure}

The singular properties of water and its essential role in our lives
have encouraged an enormous amount of theoretical and experimental
work since the first half of the twentieth century. But it has been
the search for a possible second critical point\cite{Poole1992,Franzese2001,PRL_2002_88_195701} and the careful
study of the many facets of the anomalous behaviour of water what has
been a key point for research in the last
decades\cite{PRE_1996_53_6144,PRE_2007_76_051201,JPCM_2007_19_205126,JPCM_2009_21_504108}. By
anomalous behaviour in water we mean well known features, such as the
negative slope of the p-T liquid-solid coexistence line, which in
common words means that the solid phase (ice) will melt upon
compression (see the phase diagram of Figure \ref{PDH2O}). Together with this, water is known to present a
temperature of maximum density (TMD), which at atmospheric pressure is
approximately $4^{\rm{o}}\rm{C}$, this implies that there exists a
region of anomalous behaviour in which water expands when cooled
down. At the same time one finds a region in which diffusivity
increases when pressure is increased (dynamic anomaly). When plotted
in a $T-\rho$ diagram the various anomalous regions organize in a cascade of
anomalies\cite{NAT_2001_409_318}. But not only water exhibits this
type of anomalies, other substances  as diverse as,
P\cite{PDElem1991,PRL_2003_90_255701}, Bi, C, Si, Ge, Ga, Te,\cite{PDElem1991}, silica\cite{PRE_2000_63_011202}, or
germanium oxide\cite{JCP_1995_102_6851} have 
pressure-temperature coexistence curves with negative slope, present
liquid-liquid equilibrium, or have been shown by simulation to exhibit
dynamic anomalies. All have in common
the presence of stable solid phases with relatively low coordination
numbers (3-5, lower that those of the corresponding liquid
phases).

\begin{figure}[ht]
\begin{center}
\includegraphics[width=10cm,clip]{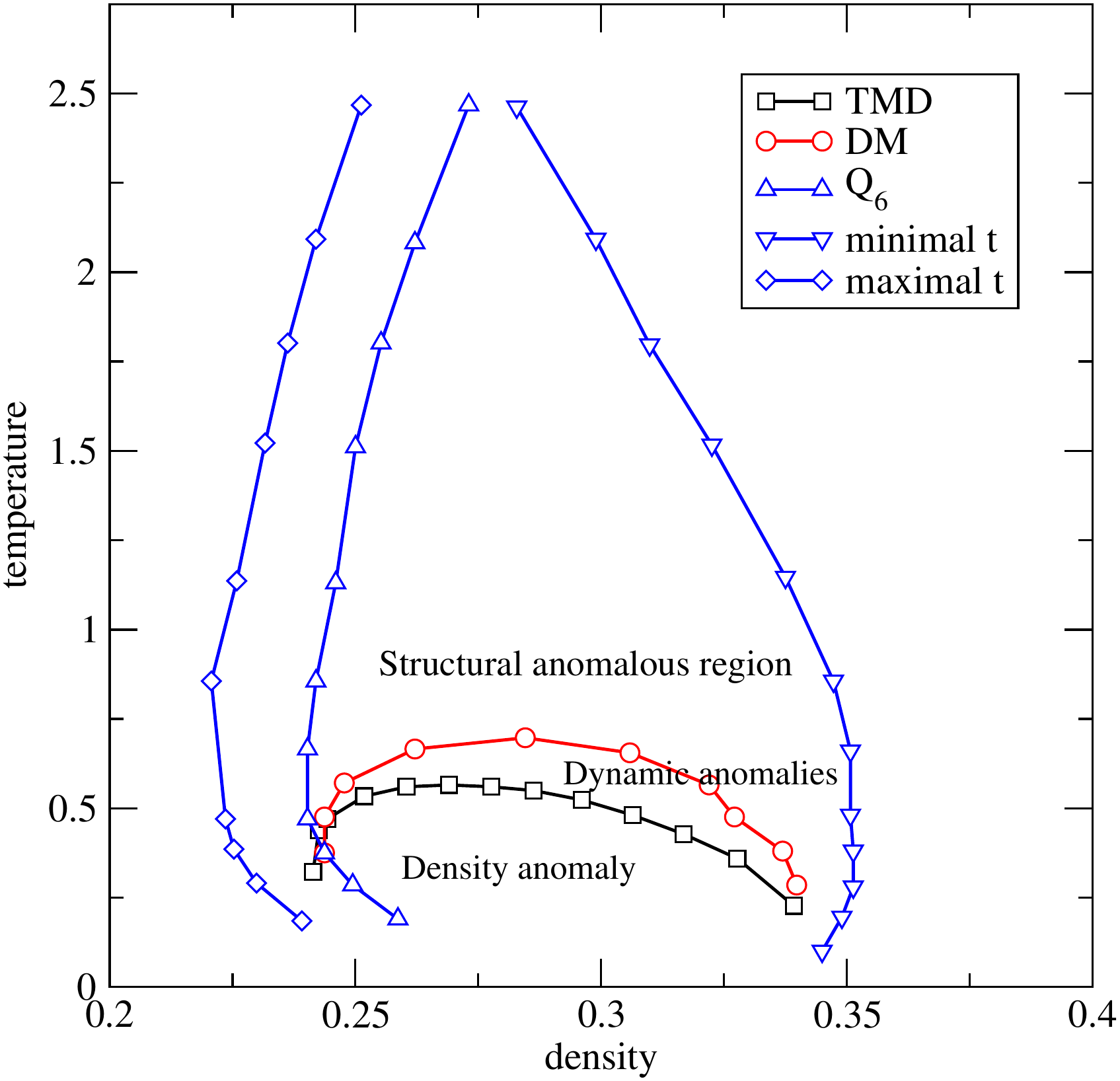}
\end{center}
\caption{Cascade of anomalies for a two-short range repulsive
  potential (Hemmer-Stell) as studied by Yan et
  al.\cite{PRL_2005_95_130604}. $Q_6$ defines the orientational order
  parameter, $t$ the translational order parameter (see
  \cite{Truskett2000}). DM designates the curve of diffusivity maxima
  and TMD the line of temperatures of density maxima.}
\label{cascade}
\end{figure}

Interestingly, it has become clear
that such anomalous behaviour can also be reproduced by model systems
interacting via spherically
symmetric potentials\cite{PRL_2005_95_130604}. In particular, the
Hemmer-Stell potential\cite{PRL_1970_24_1284}, devised with the aim of
describing simple systems that can exhibit multiple transitions, was
shown by Jagla\cite{JCP_1999_111_8980} to share with water a good
number of anomalies. In this regard, Yan
et. al\cite{PRL_2005_95_130604}, where the first to describe in this
model a cascade
of anomalies which is qualitatively similar to that of water (see
Figure \ref{cascade}). Together with the Hemmer-Stell's (or ramp) potential, a
variety of interaction models have shown to be able to display 
peculiar behaviour, such as the Lennard-Jones+Gaussian (LJG) \cite{Krott2013},
or the square-well/square shoulder (SWSS)\cite{Franzese2001} among others. It has been
argued that only the continuous 
version of the SWSS potential \cite{Salcedo2013} yields a density
anomaly, whereas other features, such as the presence of a
liquid-liquid equilibrium are reproduced by the SWSS
potential. These models are all characterized by the presence of two
repulsive ranges of interaction. There is however an additional simple
potential model that shares a good number of anomalies with these systems,
namely the Gaussian core model\cite{Mausbach2006}. Since this
potential, being finite at zero separation, 
allows for complete overlap of a pair of interacting particles, it is
a class of its own. One may think that together with a first
repulsive range defined by the width of the Gaussian,  a
second range of interaction would correspond to full overlap (or  the
maximum of the repulsive potential), thereby casting this model into
the class of two repulsive range models. 

 Additionally, some lattice gas models have also been
investigated to search for thermodynamic and structural anomalies
characterized by of both
isotropic\cite{JPCM_2004_16_8811,JPCM_2005_17_399} and 
orientational dependent
interactions\cite{JCP_2007_126_064503,PRE_2008_77_051204}. In this
contribution we will review the simplest models that can reproduce a
water-like phase behaviour and at least some of the above mentioned
anomalies. Thus we will focus on one-dimensional lattice models. In this
regard, the simplest system that can give a fluid-solid transition
(order-disorder) in one dimension is the antiferromagnetic Ising model
with two sublattices, in which the antiferromagnetic state (ordered)
is mapped into a solid in lattice gas language. The model, studied in
detail by H{\o}ye \cite{PRB_1972_6_4261} yields a second order
transition near the critical point. We will review the features of this
model in the next section. Then in Section III we will see how it can
be modified so as to reproduce a first order fluid-solid  transition
with a solid phase less dense than its fluid counterpart, following
previous work by the authors\cite{JCP_2008_129_024501}. In Section IV we will see
how the addition of a second repulsive range brings the phase diagram
closer to that of a particular class of anomalous systems such as
phosphorous\cite{MP_2009_107_321}. The 
two-parameter model can be thereafter tuned to yield a water-like phase
diagram\cite{MP_2010_108_51}. A purely repulsive version of the model
will also be shown to 
reproduce the presence of density maximum, bringing much closer the analogy to
water behaviour. The careful study of these and related simple yet
rich physical models has been a distinctive characteristic of
Prof. Johan H\o ye's scientific career, representing in many cases
relevant landmarks in Statistical Physics. 

\begin{figure}[ht]
\begin{center}
\includegraphics[width=13cm,clip]{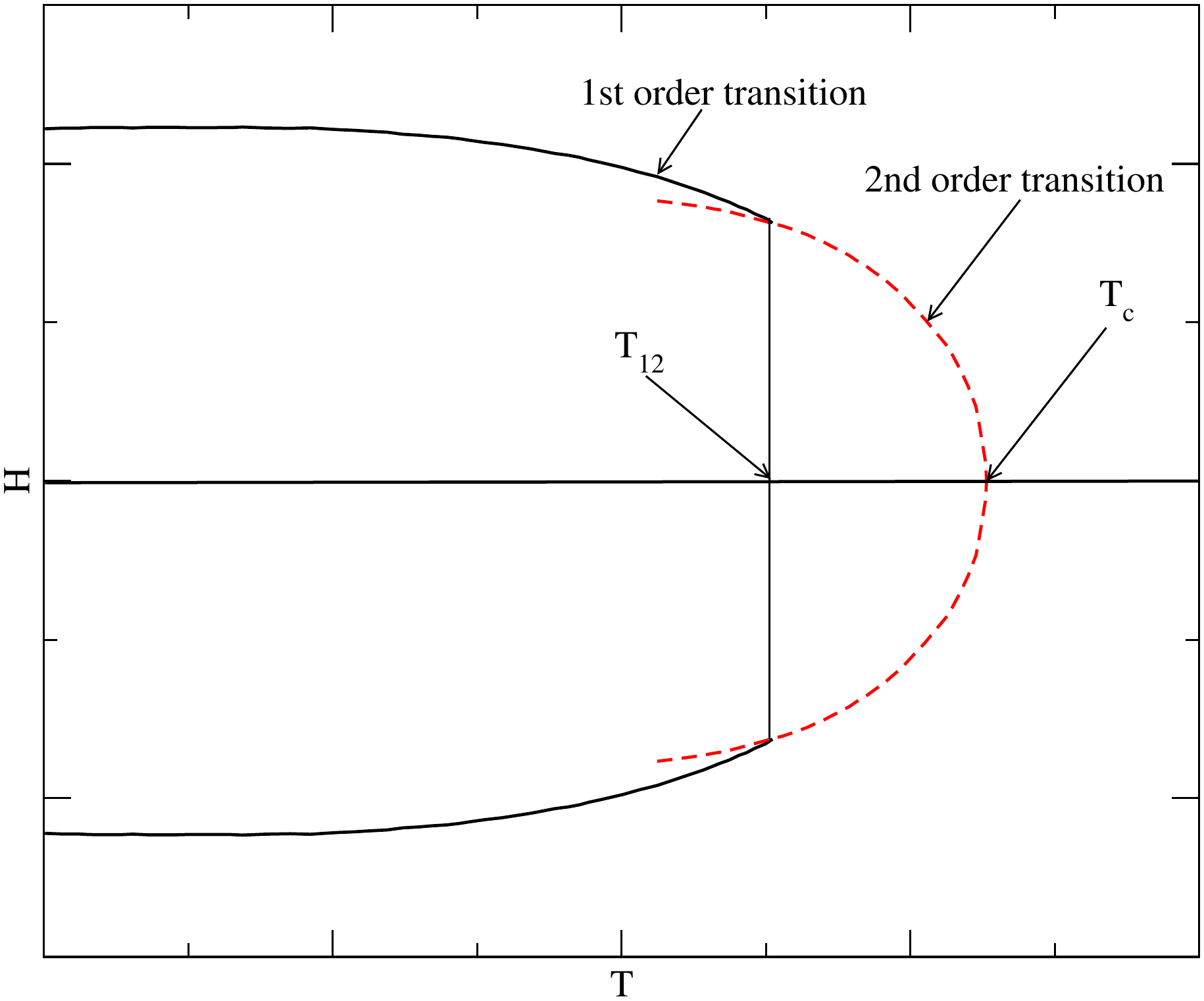}
\end{center}
\caption{Qualitative phase behaviour of the one-dimensional Ising
  model with nearest-neighbour repulsion and an infinitely long ranged
  staggered mean-field acting on even numbered neighbours\cite{PRB_1972_6_4261}.}
\label{FerroAF}
\end{figure}

\section{First step: antiferromagnetic models with two and three sublattices.} 

Back in the early seventies H\o ye \cite{PRB_1972_6_4261} studied
closely the order-disorder phase transitions of an antiferromagnetic
Ising model with two sublattices, in which the antiferromagnetic state
(ordered) can be mapped  into  a solid phase. The model
is endowed with competing short range and very long range interactions,
being the latter of Kac type, and acting only on even numbered spins,
i.e.

\begin{equation}
\varphi(i-j) =\left\{\begin{array}{cl}
-a\gamma\exp[-\gamma|i-j|], & \quad\mbox{if} \quad i-j\; \mbox{is even}\\
0, &  \quad\mbox{if} \quad i-j\; \mbox{is odd}
\end{array}
\right.
\label{kac}
\end{equation}
where $a$ is the integrated long-range interaction, i.e.
\begin{equation} a =\frac{1}{2} \sum_{n=-\infty}^\infty\varphi(i),
\label{a}
\end{equation}
and $\gamma\rightarrow 0$ being and inverse range
parameter. So this long range interaction does not necessarily
encourage ferromagnetic ordering. The full Hamiltonian can be written
as 
\begin{equation}
\mathcal{H} = J\sum_{i=1}^Ns_is_{i+1}
+\sum_{i<j}^N\varphi(i-j)s_is_j-H_0\sum_i^Ns_i.
\end{equation}
with $J>0$ being the antiferromagnetic nearest neighbour (NN) coupling
constant,  $N$ is the number of spins or lattice
sites, $s_i=\pm 1$ are the spin variables, 
$H_0$ is the external field, and we consider periodic boundary
conditions, i.e. $s_i=s_{N+i}$. This can
be shown to be one of the simplest models that can yield a fluid-solid
transition. However, in Ref. \cite{PRB_1972_6_4261} , H\o ye, showed
that such a system leads to a second order transition in a region
close to the critical point, which for temperatures below a certain
transition temperature, $T_{12}$ turns into the desired first order
fluid-solid transition (see Figure \ref{FerroAF}). 

\begin{figure}[ht]
\begin{center}
\includegraphics[width=13cm,clip]{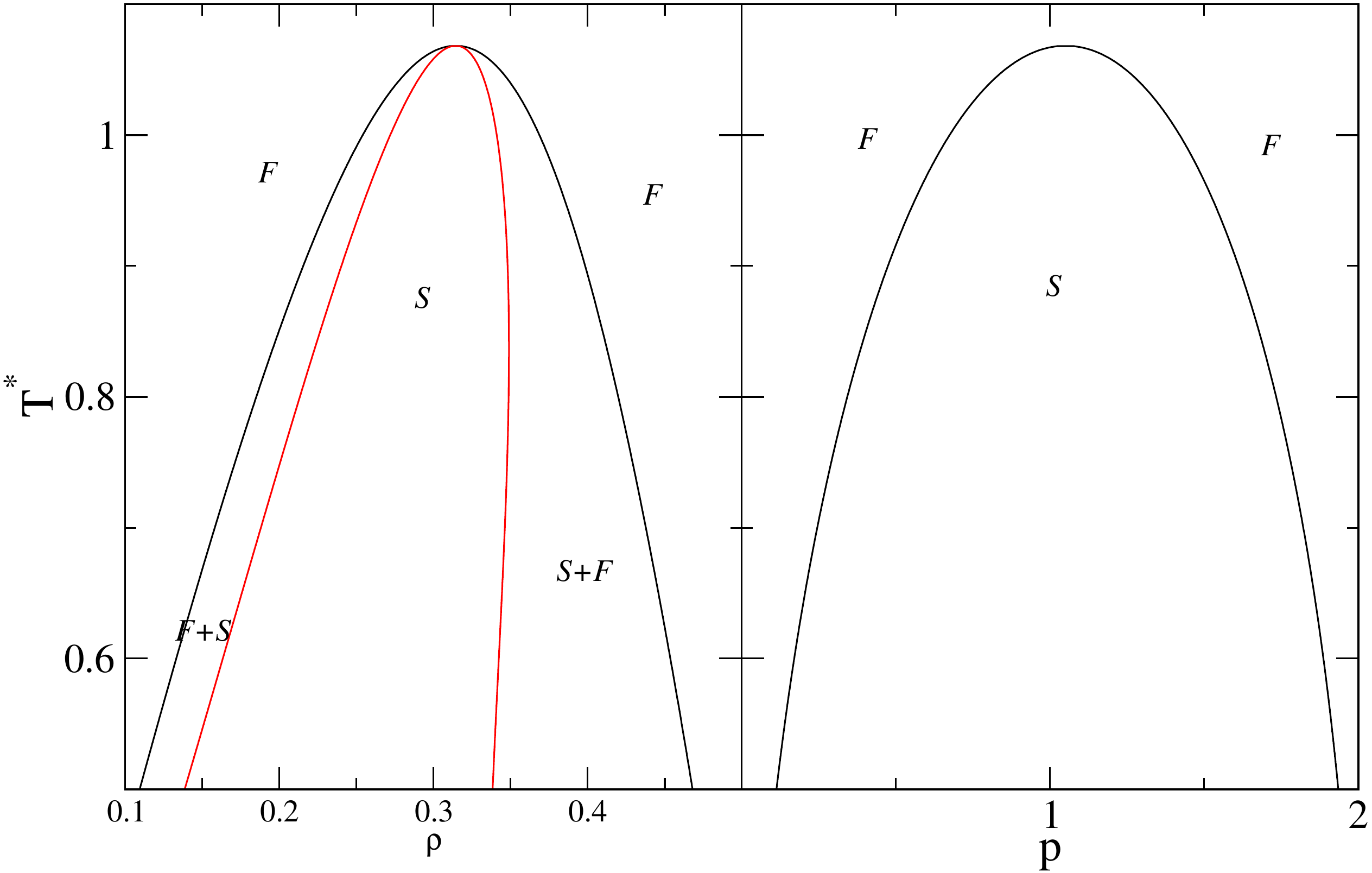}
\end{center}
\caption{Phase diagram of a lattice gas with infinite NN repulsion and
  and three sublattices with a staggered mean field.}
\label{pdSF}
\end{figure}

In order to cure this deficiency, H\o ye and Lomba
\cite{JCP_2008_129_024501}, extended the model incorporating three
sublattices, i.e. we  considered an infinitely long ranged
antiferromagnetic interaction that acts on each third spin, i.e.
\begin{equation}
\varphi(i-j) =\left\{\begin{array}{cl}
3a\gamma\exp[-\gamma|i-j|], & \quad\mbox{if} \quad i-j=3n+1,\;3n+2\\
0, &  \quad\mbox{if} \quad i-j=3n
\end{array}
\right.
\label{2}
\end{equation}
where $n$ is integer, and again the limit $\gamma\rightarrow 0$ is
considered with $a>0$. The model thus constructed would have however an
unwanted solid-solid transition in the middle of the order-disorder
transition, due to the symmetry of the Ising model when the critical
point is located at zero magnetic field or zero magnetization. The
presence of a nearest neighbour hard core  interaction will break the
symmetry with respect to the zero field, and so remove this
unwanted transition. This model can be solved using  a mean field
treatment to handle the long range staggered interaction and the
transfer matrix method to deal with the NN antiferromagnetic
coupling\cite{JCP_1971_55_4159,PRB_1972_6_4261}. As shown in \cite{JCP_2008_129_024501} one can
now get a system that yields a fluid-solid transition, in which the
solid phase melts again upon compression (see Figure \ref{pdSF}). In
order to simplify the solution, the antiferromagnetic NN coupling was
considered in the limit $J\rightarrow \infty$, by which then the
external field $H_0$ must be adjusted so that $H_0+2J$ remains finite
and spins can be turned up and down by thermal fluctuations. The model
is equivalent to a lattice gas in which the particles occupy two sites. The phase
diagram of  Figure \ref{pdSF} indicates that we
are now in a situation closer to a water-like behaviour: we have
a model in which the solid phase melts upon compression and with a
$p-T$ coexistence curve with negative slope within certain range of temperatures, i.e., pressure decreases
when temperature raises. This represents a clearly ``anomalous'' behaviour. 

\section{One step beyond: two sublattices and next-to-nearest
  neighbour interaction}

How can one improve the previous model ? On one hand we have to
account for the vapour-liquid transition. This in principle can be
achieved separating the mean field contribution to the free energy
into a staggered term (that favours antiferromagnetic long range
ordering) and an attractive term. This has to be complemented by the
incorporation of a next-to-nearest neighbour interaction (NNN), which
will enable to tune  the location of fluid-solid equilibrium
with respect to the gas-liquid critical point. 

The system  so constructed was studied by the author in collaboration
with Høye in
\cite{MP_2009_107_321}, and solved using the procedure indicated above
and which will be roughly described below. The model's Hamiltonian can
be written as 
\begin{equation}
\mathcal{H} = J\sum_{i=1}^Ns_is_{i+1}
+K\sum_{i=1}^Ns_is_{i+2}+\sum_{i<j}^N\varphi(i-j)s_is_j-H_0\sum_i^Ns_i.
\end{equation}
Again the $J > 0$ is NN coupling, and as in \cite{JCP_2008_129_024501} 
 we let $J\rightarrow\infty$ with $H_0$  adjusted
so that $H_0+2J$ remains finite, and periodic boundary conditions are
imposed, but we have now a NNN interaction with a 
coupling constant $K>0$.

\begin{figure}[t]
\centering
\includegraphics[width=13cm,clip]{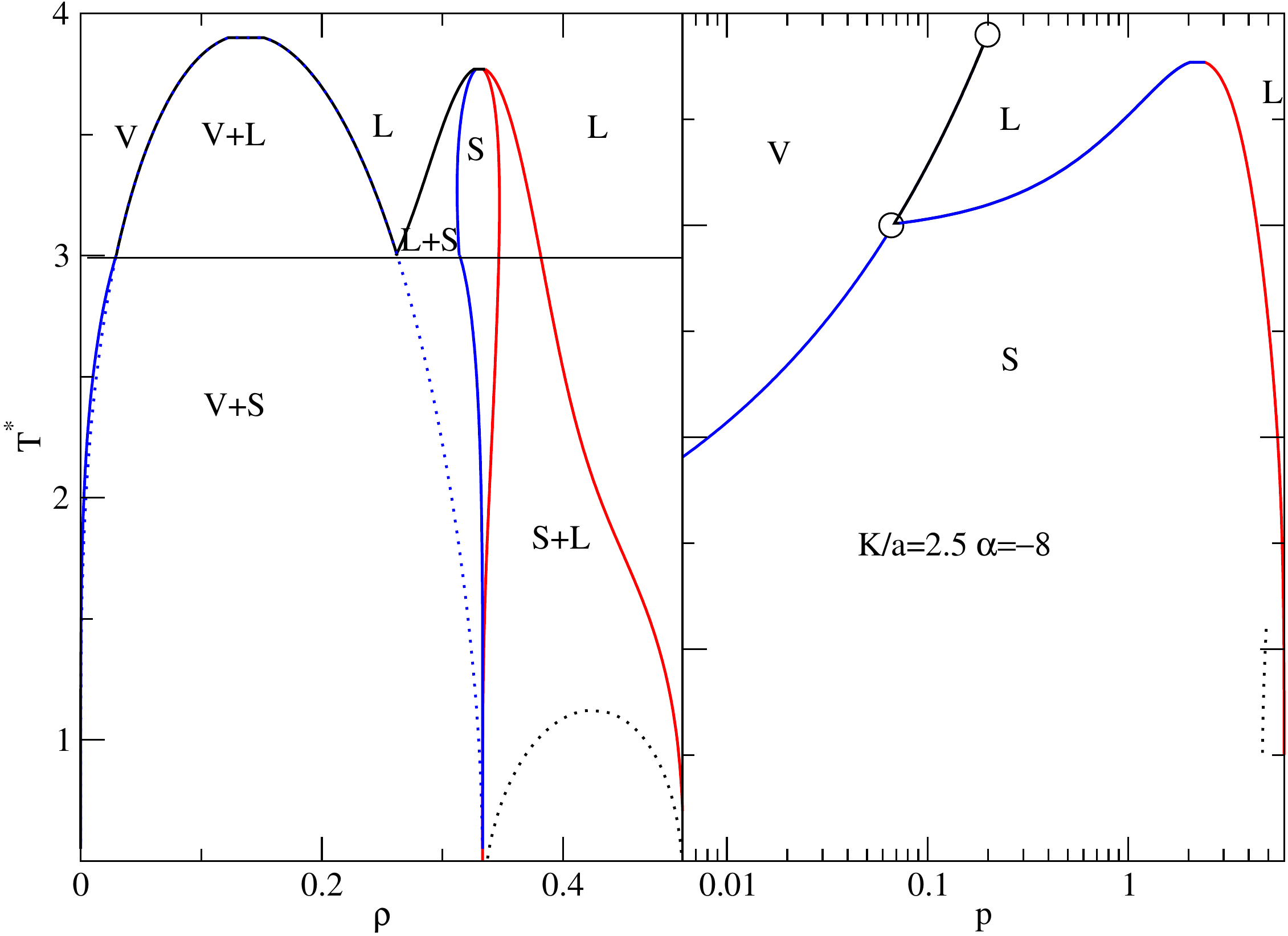}
\caption{T-$\rho$ (left) and p-T (right) phase diagrams of a one-dimensional lattice model
  with NNN interactions and a staggered field. Dotted curves indicate
  metastable equilibria.}
\label{PhasTot}
\end{figure}

The lattice can be
split into groups of cells made up of three sublattices in accordance with the
staggered interaction (\ref{2}). The effective fields, $H_i$ (acting on spins of sublattice $i$) can be
expressed in terms of the mean field interaction (\ref{2}) and the
external magnetic field as 
\begin{eqnarray}
H_1 &=& H_0 - a(m_2+m_3)\nonumber\\
H_2 &=& H_0 - a(m_1+m_3)\nonumber\\
H_3 &=& H_0 - a(m_1+m_2),
\end{eqnarray}
where $m_i$ is the magnetization per particle of sublattice $i$. Now, we
define $m_1=m+2u$ and $m_2=m_3=m-u$, where $m$ and  $u$ are the
effective and staggered magnetizations respectively. Thus the following effective field,
$H_{e}$, and effective staggered field $H_{s_e}$, can also be defined by
\begin{eqnarray}
H_{e} &=& \frac{1}{3} (H_1+H_2+H_3)=H_0-2am \nonumber\\
H_{s_e} & = & \frac{1}{3} (2H_1-H_2-H_3) ;\; u = \frac{1}{6}
(2m_1-m_2-m_3).
\label{hehse}
\end{eqnarray}
These satisfy
\begin{eqnarray}
H_0 &=& H_{e} + 2am\\
 H_{s_e} &=& 2au
\label{fields}
\end{eqnarray}
The equivalency to the lattice gas gives a number density 
 $\rho = (m+1)/2$. The reduced temperature will be
defined in terms of the mean field coupling as $T^*=1/(\beta a)$, with
$\beta = 1/k_BT$, $k_B$ the Boltzmann constant, and T the
absolute temperature as usual. The corresponding reduced field and NNN
coupling constant will be $h^*=H_0/a$ and $K^*=K/a$.

Now, in order to account for the vapour-liquid transition the effect of the uniform mean
field term $2am$ can be tuned, while retaining the staggered mean
field contribution, $2au$.  One thus ends up with
\begin{equation}
 H_{e} = H_0 - 2\alpha am
\label{healpha}
\end{equation}
where $\alpha$ will be a coupling term that can switch from uniform
mean field attraction ($\alpha < 0$) to repulsion ($\alpha>0$). 
Thus $\alpha\neq 1$ represents the additional uniform mean field interaction
mentioned below Eq.~(\ref{2}).

\begin{figure}[t]
\centering
\includegraphics[width=13cm,clip]{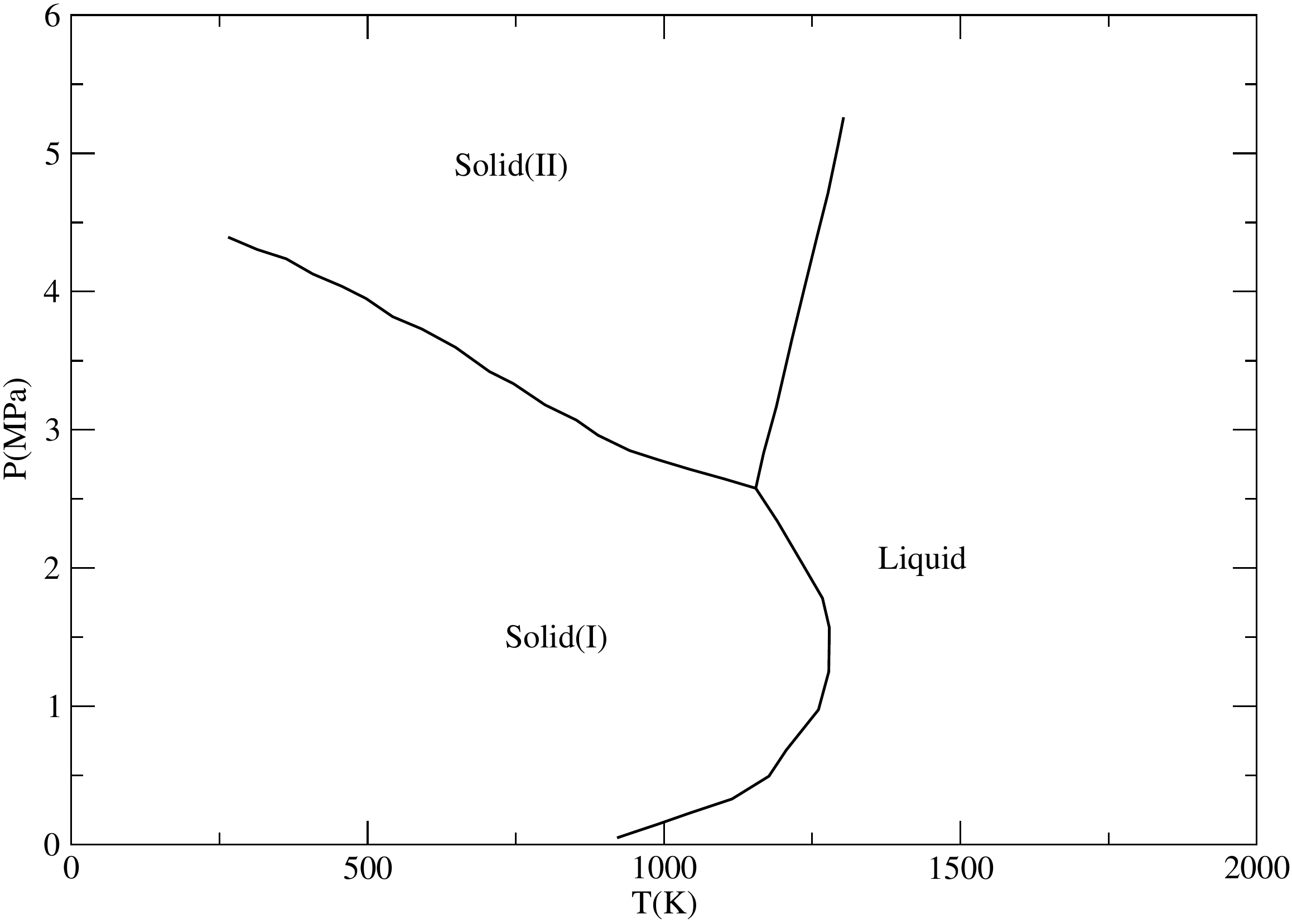}
\caption{Phase diagram of phosphorous. Data taken from Ref.\protect\cite{PDElem1991}.}
\label{PDP}
\end{figure}

The partition function, $\lambda$,
is given by the eigenvalue equation of the transfer matrix
(Eqs.~(23)-(25) in Ref.~\cite{MP_2009_107_321}), namely
\begin{equation}
\varLambda^3 + a_2\varLambda^2 +
a_1\varLambda + a_0 = 0
\label{eigen2}
\end{equation}
where $\varLambda = \lambda^3$, and
\begin{eqnarray}
a_2 & = & -[D^3p^{-3} + D(2e^{-v} + e^{2v})p] \nonumber \\
a_1 & = & -[D^2(e^{-2v} + 2e^{v})(p^{-2} - p^{2}) + p^{-6}]\nonumber  \\
a_0 & = & D^3p^{-3} (p^{-2} - p^{2})^3, \label{ai}
\label{25}
\end{eqnarray}
with
\begin{eqnarray}
D & = & \exp(-\beta H_e)\nonumber  \\
v & = & \beta H_{s_e}\;;\;p=\exp(\beta K).
\label{dv}
\end{eqnarray}

\begin{figure}[t]
\centering
\includegraphics[width=13cm,clip]{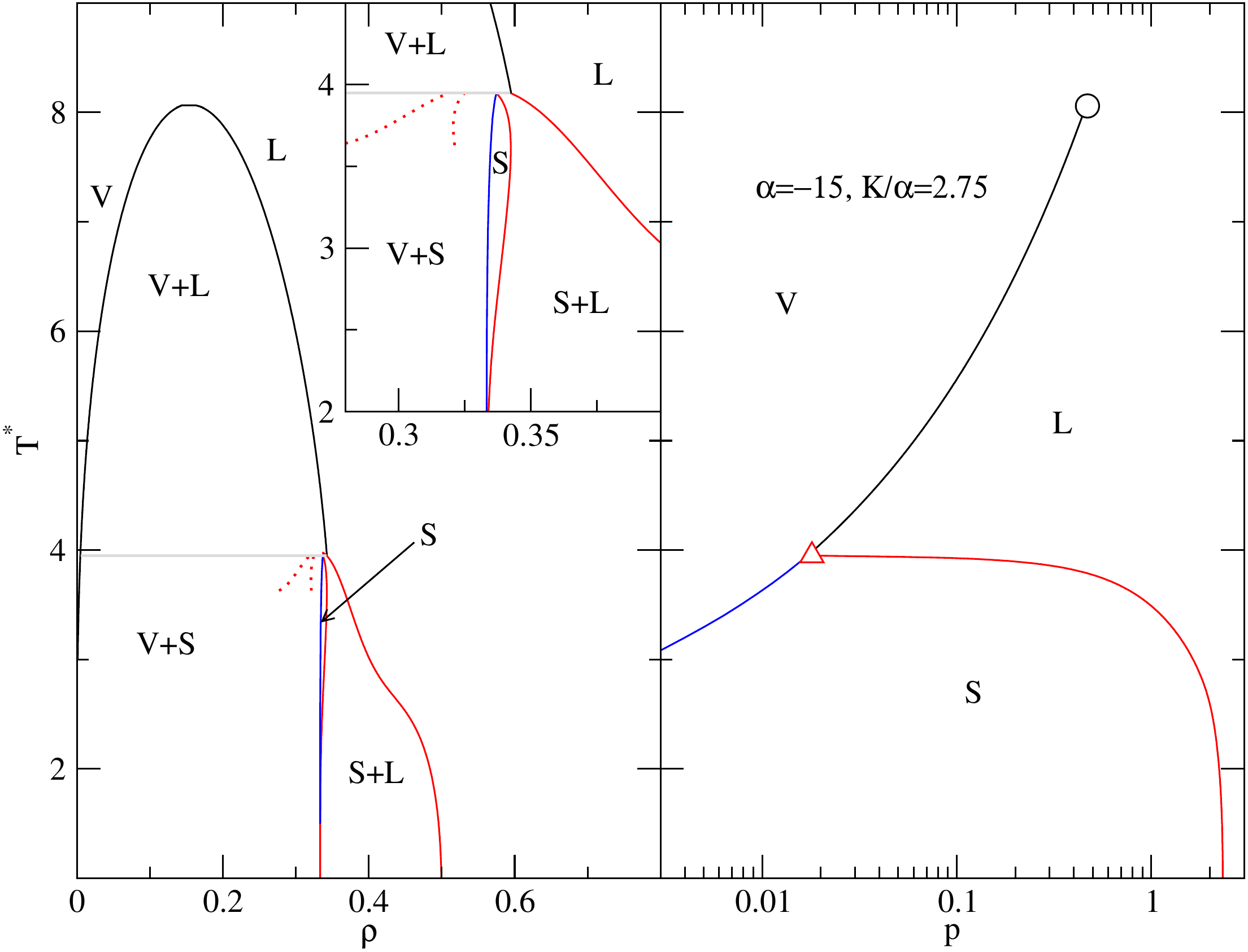}
\caption{T-$\rho$ (left) and p-T (right) phase diagrams of a one-dimensional lattice model
  with NNN interactions and a staggered field with a choice of mean
  field attraction ($\alpha$) and NNN repulsion($K/a$) to mimic the phase diagram of water.}
\label{PD}
\end{figure}

The magnetization $m$ and staggered magnetization $u$ result from the 
differentiation of the eigenvalue equation by the method used in
Ref.\cite{JCP_1971_55_4159}, provided $\lambda$ has been calculated
solving Eq.~(\ref{eigen2}). Thus, one gets
\begin{eqnarray}
m & = & -\frac{1}{3} \sum_{n=0}^3 a_n'\varLambda^n/N \label{m} \\
u & = &  -\frac{1}{3} \sum_{n=1}^3 a_{s_n}'\varLambda^n/N , \hspace{0.5cm} N = \sum_{n=1}^3
na_n\varLambda^n,
 \label{u}
\end{eqnarray}
with
\begin{eqnarray}
a_n' & = & \frac{\partial a_n}{\partial(\beta H_e)} \nonumber \\
a_{s_n}' & = & \frac{\partial a_n}{\partial(\beta H_{s_e})}, \label{an}.
\end{eqnarray}
 Explicit expressions can be found in
Ref.~\cite{MP_2009_107_321}. 

With this, the free energy $G$ per spin
\cite{JCP_2008_129_024501,MP_2009_107_321} is given by
\begin{equation}
-\beta G = \ln\lambda + \beta \alpha a m^2-\beta a u^2.
\label{betag}
\end{equation}
and the 
corresponding lattice gas pressure $p$ reads
\begin{equation}
 \beta p = -\beta G +\beta H + \beta K+\beta \alpha a.
\label{presion}
\end{equation}

We have thus all the ingredients to study
the phase behaviour of our model. To that aim, we know that the phase
equilibrium conditions determine that the field, $h^*$, and the 
spin free energy, $\beta G$, stay the same in both phases at
equilibrium. Thus one has to solve,
\begin{eqnarray}
\beta G(u=0,m_d;T^*) &= & \beta G(u_o,m_o;T^*)\label{equi1}\\
h^*(u=0,m_d;T^*) &= &  h^*(u_o,m_o;T^*)\label{eqi2}
\label{equil}
\end{eqnarray}
where the subscripts $_d$ and $_o$ denote the disordered and ordered
phases respectively. Additionally,  $m_d$ and $h^*$, and $m_o$, $u_o$
and $h^*$ are connected via 
Eqs.(\ref{m}) and (\ref{u}). Details concerning the explicit numerical
solution of these equations can be found in
Refs.~\cite{JCP_2008_129_024501} and \cite{MP_2009_107_321}.

After a careful search over the parameter space of $K/a$ and $\alpha$,
one finds that the equations (\ref{equil}) yield the phase diagram for
$K/a=2.5$ and $\alpha=-8$ as shown in Figure \ref{PhasTot}. Here both
the $T-\rho$ and $T-p$ phase diagrams are shown ($p$ is evaluated from
Eq.~(\ref{presion})). Clearly this phase diagram departs from that of
water (Figure \ref{PDH2O}) in the fact that the solid-liquid
equilibrium curve starts out from the triple point with a positive  instead of a negative
slope.  At higher densities the L-S curve reaches 
a liquidus point and then a region of anomalous behaviour
(i.e. negative slope in the $T-p$ curve) appears. As a whole, this diagram bears some resemblance to that
of phosphorous\cite{PDElem1991} depicted in Figure
\ref{PDP}. Also in Figure \ref{PhasTot} one observes the presence of a
metastable liquid-liquid equilibrium, a feature that has been argued
to be found in undercooled water (though much
debated)\cite{Poole1992,JPCM_2007_19_205126}. On the other hand, in
the case of phosphorous, a first order liquid-liquid transition has
been characterized by means of diffraction experiments in the region
of high pressures and
temperatures\cite{NAT_2000_403_170,PRL_2003_90_255701}. 

\begin{figure}[t]
\centering
\includegraphics[width=10cm,clip]{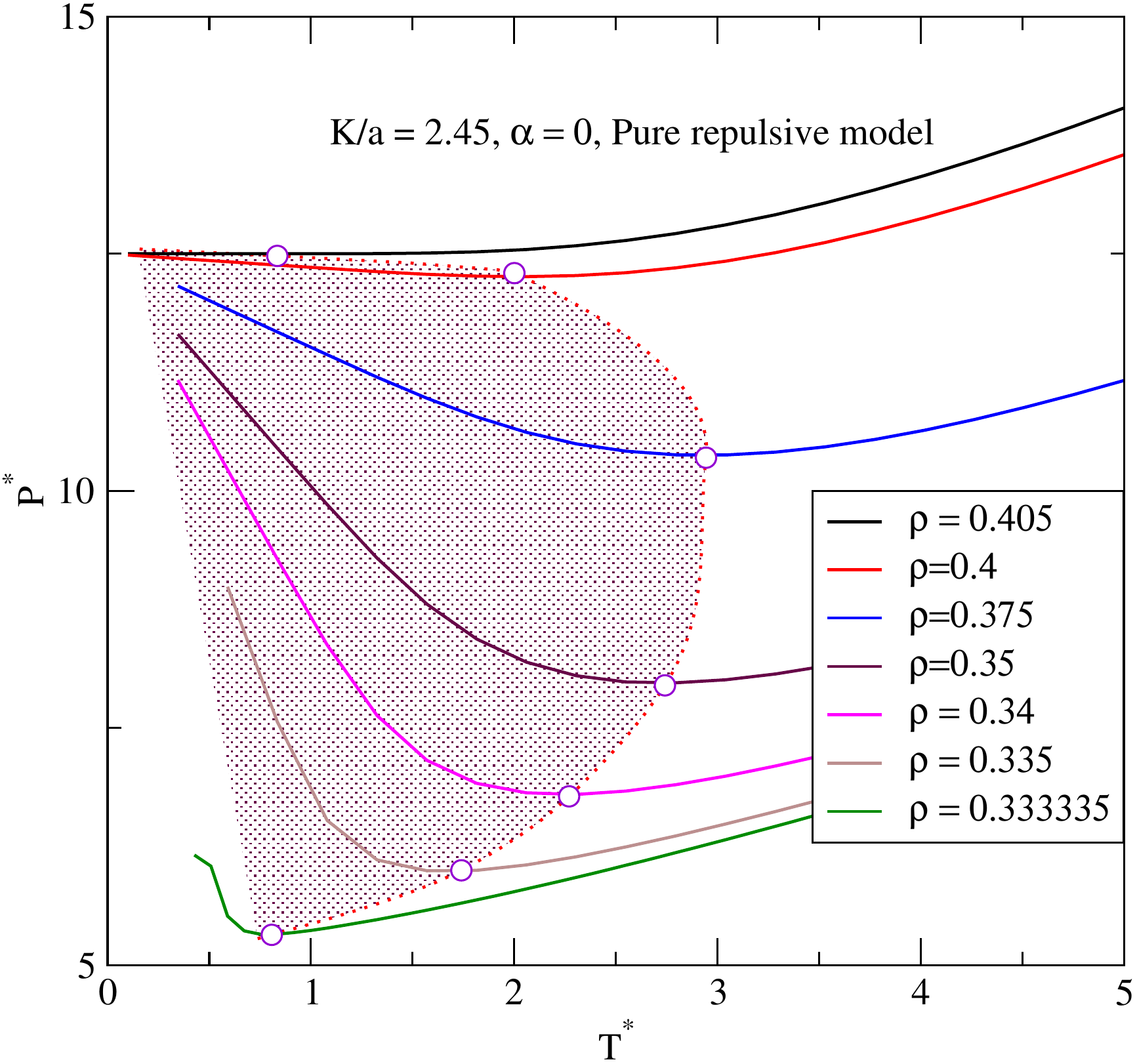}
\caption{Temperature of maximum density  curve of a one-dimensional lattice model
  with NNN interactions and a staggered field with purely repulsive
  interactions. The region of anomalous behaviour is indicated by a
  shaded area.}
\label{tmd}
\end{figure}

\section{A water-like model}

Now, the question is whether the model parameters can be tuned so as to
reproduce the behaviour of water. To that aim,  it seems
clear from Figure \ref{PhasTot} that the liquidus point has to be shifted into the vapour-liquid
equilibrium curve. The LV curve has to be made somewhat larger in that
respect so that the VL critical temperature is well above the liquidus
temperature. We found that a choice of $\alpha = -15$ (a substantial
increase in the dispersive interactions) and $K/a=2.75$ (a slight
increase in the NNN repulsion), leads to a phase diagram in
qualitative agreement with that of water, as can be seen when
comparing Figures \ref{PDH2O} and \ref{PD}. One can observe in Figure
\ref{PD} that now the LS equilibrium curve starts at the triple point
with negative slope, as expected for a water like model. The change in
density at the triple point is, on the other hand, close to one per
cent, far away from the
experimental $\Delta\rho/\rho \approx 10\%$ in real water. If one is to
bring this to a better quantitative agreement with ``real'' water,
both $\alpha$ and very specially $K/a$ must be fine tuned. This
repulsive parameter is a key quantity since being a NNN repulsion
plays an essential role in the determination of the low density solid
phases. Work on this and other refinements is currently in progress. 

We have seen then that a tuning of our two parameter model switches
from phosphorous-like  to water-like behaviour. In order to understand
the key differences between both systems, one must analyze the nature
of bonding in the different phases of water and phosphorous. In the
case of water, both the solid and the liquid phases are dominated by
the presence of relatively strong hydrogen bonds. On the other hand,
low density liquid phosphorous is a molecular liquid of P$_4$
tetrahedra which interact via weak van der Waals forces, and the solid is formed by a layered low density
orthorhombic phase\cite{PDElem1991}, which upon isothermal compression
melts back into a high density liquid with entangled chains and
clusters of covalently bonded P atoms\cite{Morishita2002}. It is
evident that attractive forces are much more important in liquid water
that in the molecular P$_4$ liquid. This
explains why in order to go from a P-like to H$_2$O-like phase diagram
one has  to substantially increase the strength of the  attractive
interaction (which rule the vapour-liquid equilibrium)  from
$\alpha=-8$ to $\alpha=-15$. As mentioned, changes in $K/a$ should make possible
the fine tuning of the density change at the triple point.

Now, a question that remains to be analyzed is whether this simple
model is able to reproduce the existence of density maxima for certain
temperatures (at constant pressure), which  is a characteristic feature of water and
related systems. Using Eq.~(\ref{presion}) to evaluate $p$ along
isochorous curves, from elementary thermodynamics we know that those
points that fulfill
\[ \left(\frac{\partial p}{\partial T^*}\right)_\rho = 0\;\mbox{also fulfill}\; \left(\frac{\partial \rho}{\partial T^*}\right)_p = 0,\]
i.e., these state points correspond to density maxima along
isobars. If one performs such a calculation with the parameters found
for our water-like model, the resulting TMD curve lies in the
metastable/unstable region. We have thus analyzed whether a corresponding purely
repulsive model lacking a LV equilibrium can capture this feature by performing the
calculations with $\alpha=0$. The result of these calculations is
plotted in Figure \ref{tmd}. As we see the curves present clear
minima which display a TMD curve that is a boundary of a
thermodynamically anomalous region, in which the fluid expands upon
cooling. This is actually the feature that was being sought. It
remains to be seen whether a sensible choice of parameters can make
compatible the existence of a TMD curve in a thermodynamically stable
region with a water-like phase diagram, thus bringing our simple model
to a much closer agreement with the experimental behaviour.

In summary, we have seen how one can construct a family of one
dimensional lattice models that can be tuned to reproduce the behaviour of
fairly complex materials such a phosphorous or water. This simple models
illustrate key physical phenomena enabling a more clear understanding
of the underlying physics of the anomalies found in the thermodynamics
of not so simple materials. They also help expanding our knowledge of
how their anomalous behaviour is originated which will in turn  help in the
development of more complex models that can reproduce more closely
the experimental behaviour.

\section*{Acknowledgment}
E.L. gratefully acknowledges the hospitality of the Institut for
Fysikk at Trondheim in which a good part of this work has been carried
out over the years, and very specially Prof. Johan S. Høye for the
very fruitful collaboration that dates back to 1987 and resulted in 17
coauthored publications. Some of these joint works are reviewed in this
contribution. Additionally, E.L. would like to acknowledge financial
support from the Direcci\'on 
General de Investigaci\'on Cient\'{\i}fica  y T\'ecnica of Spain under Grant
No. FIS2010-15502 and from the Direcci\'on General de
Universidades e Investigaci\'on de la Comunidad de Madrid under
Grant No. S2009/ESP/1691 and Program MODELICO-CM.


\end{document}